# Prediction and Measurement of the Size-dependent Stability of Fluorescence in Diamond over the Entire Nanoscale

*Carlo Bradac[1], Torsten Gaebel[1], Nishen Naidoo[1], James R. Rabeau[1,#], Amanda. S. Barnard[2,*]*

[1] Center for Quantum Science and Technology, Department of Physics, Macquarie University, Sydney, NSW, Australia, 2109.

[2] CSIRO Materials Science and Engineering, Clayton, VIC, Australia, 3168.

ABSTRACT

Fluorescent defects in non-cytotoxic diamond nanoparticles are candidates for qubits in quantum computing, optical labels in biomedical imaging and sensors in magnetometry. For each application these defects need to be optically and thermodynamically stable, and included in individual particles at suitable concentrations (singly or in large numbers). In this letter, we combine simulations, theory and experiment to provide the first comprehensive and generic prediction of the size, temperature and nitrogen-concentration dependent stability of optically active N–V defects in nanodiamonds.

[#] jrabeau@physics.mq.edu.au

[*] amanda.barnard@csiro.au

The well documented thermal, mechanical and electronic properties of the various allotropes of carbon, along with their chemical compatibility with other types of nanomaterials, make nanocarbon-based devices an attractive prospect for a variety of applications.[1] Among them, nano-sized diamond is unique, since the surfaces may be functionalized, and the lattice may contain optically active defects. These defects possess unique optical and spin properties and are already proven candidates for solid-state qubits for use in quantum information technologies.[2,3] New, emerging applications are also being identified for nanodiamonds in biotechnology and nano-medicine,[4] such as drug delivery,[5,6] spin imaging [7,8] and fluorescent biomarkers.[4,9] In the latter case, an advantage of fluorescent nanodiamond probes over alternative materials [10-12] for long-term tracking and imaging *in vivo* is, in part, based on their non-toxicity to a number of cell types.[5,13-15] In addition to this, nanodiamonds emit photons capable of penetrating tissue,[16] and exhibit a very high degree of photostability.[17-19] Each application has different nanodiamond material requirements, and rely heavily on understanding the behaviour and stability of defects in nanoscale hosts.[20]

The most widely studied defect in nanocrystalline diamond is the nitrogen–vacancy complex (N–V) [21,22] (or *color center*) which forms when a vacancy migrates to bind with a substitutional nitrogen impurity.[21-25] The energy-level structure of the negatively charged (N–V)$^-$ defect results in emissions characterized by a narrow zero-phonon line (ZPL) at 637 nm (the neutral (N–V)$^0$ center has a zero-phonon line at 575 nm) [21,26,27] accompanied by a wide structured side band of lower energy due to transition from the same excited state, but with formation of phonons localized on the defect. The optical emission from N–V centers in diamond nanocrystals synthesized by chemical vapour deposition (CVD) has been shown to strongly depend on the crystal size, and was rarely seen in small diamond nanoparticles < ~40 nm in diameter.[22] Photo-physical characteristics for 25 nm particles have been reported,[23] and most recently N–V emission from 5 nm detonation nanodiamond agglomerates[28] and isolated 8 nm diamonds [29] was shown. To date, all available data points to a strong dependence on crystal size and surface/volume ratio. In order to realise in full any of the diverse applications for N–V centers in nanodiamonds, a clearer understanding of this dependence is imperative. In particular, a predictive model which may be reliably coupled with experimental measurements would be invaluable in reaching this ambitious goal.

The need for stable fluorescent nanodiamonds at different size regimes, targeted to different applications, has prompted renewed interest in aspects such as the surface structure and reactivity, core crystallinity and the location and stability of defects in diamond. The study of impurities and defects within isolated diamond nanomaterials [30-32] and nanocrystalline diamond films [33-36] with grain sizes in the order of ~5–100 nm is receiving considerable attention. This attention has been from both experimental and theoretical perspectives, but most of the theoretical studies reported to date have concentrated on exploring the location and configuration of substitutional nitrogen in single crystal (bulk) diamond and colloidal nanodiamond, and not to the N–V complex itself.

In this paper we present a combination of computational, theoretical and experimental results examining the size dependence and stability of the N–V center in nanodiamond. To begin with, we undertook computational modeling to sample the configuration-space of $(N-V)^0$ and $(N-V)^-$ defects by substituting individual defects at over 50 geometrically unique sites along specific lattice directions in representative diamond nanoparticles containing 837 carbon atoms (~2.3 nm in diameter). These calculations were performed using the density functional based tight-binding method with self-consistent charges (SCC-DFTB),[43,44] which is a two-centre approach to density functional theory (DFT) where the Kohn-Sham density functional is expanded to second order around a reference electron density. The reference density is obtained from self-consistent density functional calculations of weakly confined neutral atoms, and the confinement potential is optimized to anticipate the charge density and effective potential in molecules and solids. A minimal valence basis is established and one- and two-centre tight-binding matrix elements are explicitly calculated within DFT. A universal short-range repulsive potential accounts for double counting terms in the Coulomb and exchange-correlation contributions, as well as the inter-nuclear repulsion, and self-consistency is included at the level of Mulliken charges, as described in reference 44. This method has the advantage of being non-periodic, and been selected for use here as it has previously been shown to be suitable for studying the crystallinity of diamond nanoparticles,[45] the electronic properties [46], and the distribution of substitutional nitrogen.[37]

The particles used in this study are a $C_{837}$ truncated octahedral bucky-diamond, and a hydrogenated $C_{837}H_{252}$ truncated octahedral nanodiamond, each displaying six {100} facets and eight {111} facets. The defects were introduced individually along specific 'substitution paths', so as to sample the full range of crystallographically and geometrically unique lattice sites within the particle, to effectively sample the configuration space. In the interest of brevity a full description of these "substitution paths" is given in reference 37, but in short: Paths A, B, C, D and E extend from the centro-symmetric lattice site out to the terminal site along directions analogous to the X, L, U, K, and W directions in the diamond Brillouin zone, respectively. In the present study, all test structures were first fully pre-relaxed using the conjugate gradient scheme to minimize the total energy, before inclusion of the N–V defects. Following inclusion of the defect, the entire structure has then been re-relaxed using the same method. In both cases, the convergence criterion for a stationary point was $10^{-4}$ a.u. $\approx 5$ meV/Å for forces.

Presented in figure 1 are the final site dependant defect energies for the (a) hydrogen passivated nanodiamond and (b) bucky-diamond. Bucky-diamond is an all-carbon core-shell particle with an $sp^3$-bonded diamond-like core, partially or completely encapsulated by an $sp^2$-bonded fullerenic/graphitic (single– or multi–layer) outer shell, while a hydrogen passivated nanodiamond is an all $sp^3$ diamond particle with each under-coordinated surface atom terminated with (a total of 252) hydrogen atoms.[38] In reality, nanodiamonds may be coated with a number of different chemical groups,[40] but hydrogen is traditionally used to test the generic effect of passivation, which is of interest here. Together, the bucky-diamond and H-terminated nanodiamond structures allow us to investigate the influence of the most important type of surface structures (being graphitized or passivated, respectively).

In figure 1, the *x*-axis represents a scaled (dimensionless) nanoparticle radius defined by dividing the distance from the center to the vacancy site $r_X$ by the total distance from the center to the extremis ($R_X$) of each path (X). Hence $r/R = 0$ is the center, and $r/R = 1$ is the outermost vacancy site located on a surface, edge or corner. Similarly, relative energy ($E_{r,X}$-$E_0$) is the total energy of the nanoparticle with a defect at $r$ along $X$ relative to the energy of the nanoparticle with the vacancy in the centro-symmetric position ($r = 0$). This figure provides an effective potential energy surface for (N–V)$^0$ and (N–V)$^-$ in diamond nanoparticles with the two dominant surface structures observed experimentally.

We are first struck by how different the results are for the bucky-diamond and the hydrogen-terminated nanodiamond, and how little the anionic charge affects the thermodynamic stability. In the case of the passivated $C_{837}H_{252}$ structure (figure 1a and 1c), (N–V)$^0$ defects are relatively stable within the particle until the substitution site is within 3 atomic layers from the surface/edge/corner. The small energy fluctuations for r/R < 0.6 are due to the redistribution of the excess charge from the donor electron that occur when the defect site is within the Bohr excitonic radius of the particle extremes [37]. In this region, the nitrogen atom is sp$^3$ hybridized and the configuration of the defect is constrained. Although the defect is thermodynamically unfavorable, the energetic barrier for a transformation to a lower energy configuration is too high. At r/R >0.7 there is a ~1.5 – 4.5 eV thermodynamic driving force for diffusion that increases the closer the vacancy is to the surface/edge/corner.

In the $C_{837}$ bucky-diamond structure (figure 1b and 1d) the defect is highly unstable, and with a substantial thermodynamic (up to ~6 – 9 eV) driving force for diffusion within the particle core. In the bucky-diamond particles, where the sp$^2$-shell exists, the lattice parameter of the particle is different from the bulk, and a significant amount of strain already exists in the particle. This energy barrier for distortion of the (N–V)$^0$ and (N–V)$^-$ defects is lowered, and depending on the position of the defect the structure of the defective region changes to reduce the total stress and the total energy. This manifests as sub-surface graphitization, since this the defect energy for (N–V)$^0$ and (N–V)$^-$ is lower in sp$^2$-bonded regions, and a significant reduction in the site-dependent defect energy. These transformations are common near the surface of nanodiamonds, even when hydrogen terminated, and can be identified in the cores of bucky-diamonds by the large energy fluctuations in figures 1b and 1d which appear as a series of local minima. More information on this is provided in references 37.

Based on these results, one may conclude that both (N–V)$^0$ and (N–V)$^-$ defects will be stable deep within the core of diamond nanoparticles [39], but will only stay that way when the particles are sufficiently large so as to be predominantly bulk-diamond-like, or they have stable passivated surfaces. In the proceeding analysis, results for the (N–V)$^-$ are used, denoted simply as N–V, but the results for the neutral (N–V)$^0$ defect have been calculated and are equivalent to (N–V)$^-$ within the accuracy of the theoretical and computational methods applied. This is intuitively obvious when one considers the similarity of the raw data (figure 1), upon which the following analysis is based.

It is also possible to use these results to predict the effective concentration of defects and the probability of observation based on the kinetic barriers. The probability of observation ($P_{obs}(R,E_K)$) of a N–V defect in a diamond nanoparticle of radius ($R$) is a function of the kinetic energy during probing $E_K$, the probability that the defect will diffuse to the surface and escape ($P_{esc}(R,E_K)$) and the probability that a N–V defect will be initially created during synthesis ($P_{form}(R)$), such that:

$$P_{obs}(R,E_K) = P_{form}(R)[1 - P_{esc}(R,E_K)] \qquad (1)$$

The probability of the formation of a N–V defect will be proportional to the concentration of nitrogen present during synthesis ($C$), the kinetic energy during growth ($E_{K,growth}$), and will be a function of the characteristic energy of the defect $E_d$ at a position $r$. This may be approximated by a Boltzmann function, so that:

$$P_{form}(R) = C \sum_{r=0}^{R} P(r) \exp\left(-E_d(r) / E_{K,growth}\right), \qquad (2)$$

where $P(r)$ is the probability of the defect being at $r$, when $0 < r < R$. We can see from the results above, that there are two distinct structural environments which may surround an N–V defect. It may be in a $sp^3$-bonded environment, in the bulk-like *core* region, or in a $sp^2$-bonded environment, such as in the shell or when the defect induces localized sub-surface graphitization. We may therefore simplify this to:

$$P_{form}(R) = C\left[P_{core}(R_{core}) \exp\left(-E_{d,core}/E_{K,growth}\right) + P_{shell}(R - R_{core}) \exp\left(-E_{d,shell}/E_{K,growth}\right)\right], \qquad (3)$$

where $E_{d,core}$ is the characteristic energy of the defect in the $sp^3$-bonded core region, $E_{d,shell}$ is the characteristic energy of the defect in the $sp^2$-bonded shell region, $R_{core}$ is the radius of the core, and $P_{core}$ and $P_{shell}$ are the probability of the defect being located in the core and shell, respectively. In the case of the former terms, it is possible that coating the surface of the particle with oxygen (or another molecule) would give slightly different computational results in the shell, but it is unlikely that the extent of the shell region would be altered as this has been shown to be related to the excitonic radius of the nitrogen, and not to the specific surface chemistry.[37] In addition to this, the configuration of the defect is different in the shell, and is not likely optically active, as it is in the core. The latter are related to the fraction of atoms occupying each region, so that we may use $P_{core} = N_{core}/N$ and $P_{shell} = 1 - N_{core}/N$. Shenderova et al.[41] determined that the total number of atoms ($N$) in a cuboctahedral diamond particle with $n$ atoms along the (111)/(111) edge is given by:

$$N = \begin{cases} \dfrac{1}{12} n(2n+1)(5n+2) & \forall n \in (2,4,6,8...) \\ \dfrac{1}{12}(10n^3 + 9n^2 + 2n - 9) & \forall n \in (1,3,5,7...) \end{cases} \quad (4)$$

Since we know from the energy calculations presented above that the shell region consists of 4 to 8 atomic layers for H-terminated nanodiamond and unpassivated bucky-diamond respectively (see figure 1), we can also use this formula to determine $N_{core}$ by simply calculating the number of atoms in a particle that is the size of the core. Note that if the nanoparticle has less than 5 atoms along the (111)/(111) edge, then it is effectively "all-shell", as indicated in reference 37.

Similarly, the total probability of escape will be a combination of contributions from the core and shell regions. Each probability of escape will be a function of the input kinetic energy ($E_K$), due to a combination of the probe and the environment, and the escape energies. These escape energies are denoted by $E_{esc,core}(E_K)$ and $E_{esc,shell}(E_K)$ for the core and shell, respectively, and are once again described using Boltzmann function. If the $E_K$ of the probe is significantly lower than the escape energy for each region then $P_{esc}(E_K)$ will be negligible, whereas when $E_K = E_{esc}(E_K)$ the probability for diffusion approaches unity in that region. Hence, the total probability of escape, $P_{esc}(R, E_K)$, is:

$$P_{esc}(R,E_K) = [P_{core}(R)P_{esc,core}(E_K) + P_{shell}(R)P_{esc,shell}(E_K)]$$
$$= \left[\frac{N_{core}}{N}\exp\left(-E_{esc,core}/E_K\right) + \frac{N-N_{core}}{N}\right]\exp\left(-E_{esc,shell}/E_K\right), \quad (5)$$

where $E_{esc,core} = |E_{diff,core} - E_{d,core}|$ and $E_{esc,shell} = |E_{diff,shell} - E_{d,shell}|$ are the differences in the kinetic barrier to diffusion $E_{diff}$ and the energy of the static defect $E_d$, in the core and shell, respectively. Therefore, the calculation of $P_{obs}(R,E_K)$ requires only $C$, $E_{d,core}$, $E_{d,shell}$, $E_{diff,core}$ and $E_{diff,shell}$.

The average values of $E_{d,core}$ and $E_{d,shell}$ have been calculated explicitly for nanodiamond. We find that $E_{d,core} = 5.17\pm0.2$ eV, and $E_{d,shell}$ is $0.60\pm0.2$ eV in the $C_{837}$ bucky-diamond and $3.81\pm0.2$ eV in the $C_{837}H_{252}$ H-terminated nanodiamond. These values are obtained from the same data sets displayed in figure 1, where $E_{d,core}$ has been subtracted to obtain the relative particle energies in each case. The diffusion of N–V defects is vacancy assisted, and is dominated by the N–C exchange energy. Given that the $sp^3$-bonded core is defined as being bulk diamond-like, we have calculated the diffusion barrier for a neutral N–V defect in bulk diamond to be $E_{diff,core} = 6.68\pm0.2$ eV, which is in good agreement with previously reported values.[24,42] Similarly, as we have defined the $sp^2$–bonded shell to be graphitic, we have calculated the ($c$-axis) diffusion barrier for a neutral N–V defect bulk graphite to be $E_{diff,shell} = 13.3\pm0.2$ eV.

Using $E_K = 800$ ºC and $C = 0.1\%$ from experiment,[22] an estimate of the probability of observation of a N–V defect in H-terminated nanodiamond is shown in figure 2a. We can see that particles $< \sim 22$ nm in diameter have a $< 1\ \%$ probability of containing a stable N–V defect, even under ambient conditions. Alternatively, equation 1 is also dependent on the synthesis temperature ($E_{K,growth}$) and $C$, as we can see from equation (2), so a complementary plot representing the formation conditions may also be generated. This is contained in figure 2b, where the probability of observation as a function of the synthesis temperature and the nitrogen concentration in the precursor materials is given. As one would expect, increasing the synthesis temperature and quantity of N in the precursors increases the probability that N–V defects will be present in the lattice. Naturally, these estimations assume that diffusion occurs on the same timescale as observed in bulk diamond and graphite.

Together the computation and theoretical results indicate that the stability of N–V centers is greater within the core of nanodiamonds, and hence the probability of observation is greater for larger particles (where a greater number of lattice sites occupy the core region). They also show how the growth temperature and N concentration affect the incorporation of N–V centers, and provide a predictive framework in 4-dimensional $<D,E_K,E_{growth},C>$ space. A 4-D comparison of this type cannot be achieved via exclusively computational or experimental techniques with undertaking a very large number of individual investigations and then mapping the results onto the manifold. As mentioned above, there are three distinct synthesis methods used to produce diamond nanoparticles, each resulting in characteristic sizes, morphologies and end uses. To see how this model relates to different diamond nanoparticles produced with different synthesis techniques, we refer to figure 3, where results are predicted for HPHT nanodiamond, CVD nanodiamond, and the highly desirable detonation or ultra-dispersed diamond (UDD) nanoparticles.

Experimental measurements have been conducted to determine the relationship between the size of a diamond nanocrystal and the probability of finding optically active N–V center(s) in them. As mentioned above, observations of the efficiency of N–V in nanocrystalline diamond films produced using the CVD method have previously been reported for a sample with ~20 nanocrystals exhibiting characteristic fluorescence.[22] In this case the probability of N–V defects was found to decrease from around ~15 % in grains ~100 nm in diameter to around 2 % for grains between 60-70 nm in diameter. Results corresponding to H-terminated CVD nanodiamond observed at room temperature are shown in Figure 2a, where we predict a probability of between 2 – 3 % for N–V stability in particles between 60 – 70 nm in diameter. This is in good agreement with the previously reported experimental results.[22] Alternatively, using the computational results in figure 1 and diffusion barriers for graphitized bucky-diamond particles in equations (3) and (5) gives a prediction of the probability of N–V defects being stable in 5 nm UDD nanodiamonds at room temperature (with 300ppm of nitrogen) of 0.0017 %, and only 0.00004 % in 3 nm particles. While this is incredibly low, the theoretical prediction is in good agreement with the experimentally measured value of 0.00015 % reported in 28. The variation in the predicted values for UDD are due to the particle, but may alternatively be introduced if we assume a monodisperse sample and allow for variations in temperature or nitrogen concentration. In the case of a 4 nm particle, a probability distribution of 0.0006 ± 0.0002% can equally be due to a difference in the probe kinetic energy of ±0.1 eV or a difference in the source nitrogen concentration of ±100ppm. Using the model herein it is a simple matter to assess the uncertainties associated with the natural distributions in real samples, however a statistically robust comparison with experiment is required before this model can be trusted as a generic predictor of N–V stability in nanodiamonds, in particular focusing on the HPHT nanodiamonds which have not yet been explicitly measured in this context.

Measurements of HPHT nanodiamond was carried out by analyzing a sample region of $50 \times 50$ μm, containing a total of 3690 diamond nanocrystals (~1.5 crystals per μm$^2$), 94 of which exhibited N–V fluorescence. Figure 4 shows the dataset from the analyzed sample, collected with the combined confocal/atomic force microscope (AFM) system (figure 4a). The sample we studied consisted of monocrystalline diamond powder (Microdiamant, MSY 0-0.1 micron) dispersed on a 170 μm thick glass coverslip (Menzer-Glaser). The sample was characterized with a lab-built, room temperature confocal sample-scanning fluorescence microscope (100× oil immersion objective lens, NA 1.4) combined with a commercial AFM system (NT-MDT) as similarly described in reference 22. Excitation of the N–V centers was achieved with a 532 nm CW diode pumped solid-state laser (Coherent, Compass). A notch and long pass filter were used in detection to cut off the pump laser beam and to measure the red-shifted fluorescence of the N–V centers, as shown in figure 4a.

The laser of the confocal system and the cantilever tip of the AFM were aligned to be coincident on the sample. This alignment allowed direct comparison of the fluorescing N–V centers and the host crystal in which the N–V's themselves were incorporated. Identification of the N–V fluorescence was measured using a commercial spectrum analyzer (Princeton Instruments, Acton 2500i). A Hanbury Brown and Twiss (HBT) interferometer was then used to determine if the detected fluorescence was due to single or multiple N-V centers by measuring the second order correlation function ($g^{(2)}(\tau) = \langle I(t)I(t+\tau)/I(t)^2\rangle$). $g^{(2)}$ is the probability of detecting two simultaneous photons ($\tau = 0$) normalized by the probability of detecting two photons at once for a random photon source: an "antibunching" dip in $g^{(2)}$ indicates sub-Poissonian statistics of the emitted photons and reveals the presence of single quantum-system which cannot simultaneously emit two photons. The contrast in $g^{(2)}$ approximately scales as $1/N$, where $N$ is the number of emitters. The precise size of a crystal is therefore matched with the number of N–V centers present inside that crystal.

In figure 4b we see the confocal fluorescence data, where the spots indicate fluorescing centers. Figure 4c is the corresponding AFM data, where the spots identify the heights of the crystals on the sample surface. Note that in (b) the brighter regions correspond to a greater fluorescence intensity, while for (c) the brighter regions correspond to a larger vertical dimension of the corresponding crystal. Figure 5a shows the raw coincidence $c(t)$ (right axis) and the autocorrelation function $g^{(2)}(t)$ (left axis) for three representative crystals in the sample. The raw coincidence rate $c(t)$ counted during a time $T$ within a time bin of width $w$ is normalized to that of a Poissonian source according to the formula $C_N(\tau) = c(\tau)/N_1 N_2 w T$, where $N_{1,2}$ are the count rates on each detector of the HBT interferometer. The normalized coincidence rate $C_N(\tau)$ is corrected for the background light to obtain the autocorrelation function $g^{(2)}(t) = [C_N(\tau)-(1-\rho^2)]/\rho^2$, where $\rho = S/(S+B)$ is related to the signal ($S$) to background ($B$) ratio measured independently for each emitter.[47,48] Curves A and B correspond to crystals containing more than one emitter ($g^{(2)}(0) > 0.5$), curve C shows $g^{(2)}(0) \approx 0$ indicating a single N–V center; the slight offset from zero is attributed to the remaining background fluorescence of the quartz substrate and of the diamond nanocrystal hosting the colour center.

Based on this analysis, Figure 5b is a histogram of the size distribution measured with the AFM for the 3690 nanodiamond crystals on the glass surface, and figure 5c shows the histogram of the probability of finding N–V colour centers in a range of crystal sizes.[49] Figure 5c indicates that the probability of finding N–V centers in HPHT nanodiamond is negligible when the crystal diameter is < 30 – 35 nm. The probability of finding 1 N–V incorporated in a diamond nanocrystal (figure 5c, red) becomes significant for crystals of ~ 35 nm in diameter, and a probability of finding more than 1 N–V (figure 5c, grey) becomes significant at ~55 nm in diameter. These thresholds assume the volume scales approximately exponentially with the measured diameter, and that the vacancies have a higher probability to annihilate when they are closer to the surface.[14] These experimental observations justify the assumptions made in equation (1), where $P_{obs}(R,E_K)$ is related to the probability $P_{form}(R)$ of having an N–V defect initially formed, and the probability $P_{esc}(R,E_K)$ that the N–V defect (or associated vacancy) diffuses to the surface and annihilates.

Both $P_{form}(R)$ and $P_{esc}(R,E_K)$ are a function of the radius $R$ of the particle, which can be directly related to the diameter of the crystals measured experimentally and shown in figure 5b and 5c. Our experimental observations suggest that for the sample we characterized, a radius $R$ = 15 nm (approximately half of the measured height of 30 – 35 nm, assuming a simple spherical shape for the host crystal with the N–V center in the middle) is required to avoid the vacancy diffusion-annihilation at the surface ($P_{esc}(R,E_K)$ = 1). From the histograms of figure 5 the probability of finding N–V increases non-linearly with crystal size, in agreement with theoretical probabilities of finding 1 N–V in nanoparticles of a certain size (see figure 2a). Note that beyond 100 nm in diameter, the number of crystals present was too small to be statistically significant (figure 5). The experimental measurements and theoretical predictions are directly compared, as shown in figure 5d, where we find they are in excellent agreement.

We can see that the trend in experimentally measured N–V content agrees with the theoretically predicted trend for H-terminated or bucky-diamond (partially graphitized surface), or some weighted combination of the two. Although our samples were cleaned using strong oxidizing acids and therefore the surfaces were oxygen terminated, as discussed previously, the difference between an O or H terminated diamond surface is essentially irrelevant until the defect is very close to the surface (where it has already been shown that N–V is not stable). Along with the predictions of the stability and observation of photoactive N–V centers in nanocrystalline diamond, this information is extremely useful for those wishing to customize nanodiamond samples for use in quantum or bio-medical applications [20].

We have successfully predicted the probability of measuring stable $(N–V)^0$ center(s) or $(N–V)^-$ center(s), since they have been found to be thermodynamically equivalent in diamond nanocrystals over the entire nanoscale; and have used computational results, a theoretical predictive model and direct experimental analysis to clearly demonstrate the correlation between the probability and the size of the host diamond nanoparticles. Both the theoretical and the experimental approaches show a non-linear increasing probability of finding either N–V centers as the dimensions of the crystal increase. The increasing trend varies depending upon the synthesis methods (characterized by the temperature and nitrogen concentration during growth) and the type of surface terminations. A direct comparison of the size-dependent probability for HPHT, CVD and UDD sample with a probe (kinetic) energy of 532 nm is shown in figure 6, where these differences are immediately apparent. In general, the results for the hydrogen terminated nanodiamond particles scale as $1/R^3$, whereas the bucky-diamonds scale as $1/R^{2.72}$. This is, of course, a consequence of the defects being stable only within the core region of the bucky-diamonds, which has a reduced radius with respect to the total size of the particle. Our analysis is complementary to the previously reported statistical Monte Carlo results (rather than competitive), as it considers the probability of observation of an N-V in given particle volume, as opposed to the probability of diffusion of a vacancy via a random walk through the diamond lattice.[28] Unlike the previous study however, the present approach allows for different surface structures, growth temperatures and probe energies to be considered for the very first time.

Furthermore, the experimental measurements provide additional identification of the critical dimension for which (under the conditions employed), the probability of finding a single N–V defect is optimal. We have demonstrated experimentally and in theory that irrespective of the method of incorporation, N–V defects are significantly more stable in nanodiamond matrices with dimensions sufficiently large to show at least *some* bulk-diamond behaviours. It has been demonstrated experimentally that ion-implantation can enhance the occurrence of N–V in nanodiamonds relative to as-grown material. In this paper we have described cases where N is incorporated during growth, but our method could also be applied to ion bombardment by replacing equation (3) with an appropriate term, or extending the model to include this step.

In addition to the key results, the analysis presented in this paper shows how theoretical and experimental approaches can work effectively together to improve our understanding of diamond as a host to scientifically and technologically important defects. We have generated and verified a predictive framework for those working in different application spaces wishing to understand the baseline properties of the available material. From this point, there exists considerable scope for expanding the model to include a range of processing conditions. A large portion of the literature over the past few years, which details the considerable promise of N–V nanodiamonds in biological or physical technologies, points to a similar road-block faced by everyone: the challenge of controlling and optimizing the available material. For instance, the recent ground breaking results in nanodiamond based magentometry are highly dependent on material size and quality. This work takes an enormous leap forward in meeting these challenges by strongly coupling an experimental and theoretical approach.

ACKNOWLEDGMENT: The authors would like to acknowledge support from the Australian Research Council Discovery Projects and Fellowship (JRR), Macquarie University Research Innovation Fund (JRR, NN), the Macquarie University Research Scholarships (CB). Computational resources were provided by the Oxford Supercomputing Center, the Victorian Partnership for Advanced Computing (VPAC) and the Australian National Computing Infrastructure (NCI) (ASB).

REFERENCES
1. Shenderova, O.A.; Zirnov, V.V.; Brenner, D.W. *Crit. Rev. Solid State Mater*. Sci. **2002**, *27*, 227–356.

2.  Wrachtrup, J.; Jelezko, F. *J. Phys.: Condens. Matter* **2006**, *18*, S807–S824.

3.  Gurudev Dutt, M. V.; Childress, L.; Jiang, L.; Togan, E.; Maze, J.; Jelezko, F.; Zibrov, A. S.; Hemmer, P. R.; Lukin, M. D. *Science,* **2007**, *316*, 1312–1316.

4.  Chao, J.-I.; Perevedentseva, E.; Chung, P.-H.; Liu, K.-K.; Cheng, C.-Y.; Chang, C.-C.; Cheng, C.-L. *Biophys. J*. **2007**, *93*, 2199–2208.

5.  Huang, H. H.; Pierstorff, E.; Ōsawa, E.; Ho, D. *Nano Lett.* **2007**, *7*, 3305–3314.

6.  Lam, R.; Chen, M.; Pierstorff, E.; Huang, H.; Ōsawa, E.; Ho, D. *ACS Nano*, **2008**, *2*, 2095–2102.

7.  Balasubramanian, G.; Chan, I. Y.; Kolesov, R.; Al-Hmoud, M.; Tisler, J.; Shin, C.; Kim, C.; Wojcik, A.; Hemmer, P. R.; Krueger, A.; Hanke, T.; Leitenstorfer, A.; Bratschitsch, R.; Jelezko, F.; Wrachtrup, J. *Nature,* **2008**, *455*, 648–651.

8.  Maze, J. R.; Stanwix, P. L.; Hodges, J. S.; Hong, S.; Taylor, J. M.; Cappellaro, P.; Jiang, L.; Gurudev Dutt, M. V.; E. Togan, Zibrov, A. S.; Yacoby, A.; Walsworth, R. L.; Lukin, M. D. *Nature*, **2008**, *455*, 644–647.

9.  Yu, S.-J.; Kang, M.-W.; Chang, H.-C.; Chen, K.-M.; Yu, Y.-C. *J. Am. Chem. Soc.* **2005**, *127*, 17604–17605.

10. Medintz, I. L.; Uyeda, H. T.; Goldman, E. R.; Mattoussi, H. *Nature Mater*. **2005**, *4*, 435.

11. Cui, B.; Wu, C.; Chen, L.; Ramirez, A.; Bearer, E. L.; Li, W.-P.; Mobley, W. C.; Chu, S. *Proc. Natl. Acad. Sci. USA*, **2007**, *104*, 13666–13671.

12. Akin, D.; Sturgis, J.; Ragheb, K.; Sherman, D.; Burkholder, K.; Robinson, J. P.; Bhunia, A. K.; Mohammed, S.; Bashir, R. *Nature Nanotech*. **2007**, *2*, 441–449.

13. Liu, K.-K.; Cheng, C.-L.; Chang, C.-C.; Chao, J.-L. *Nanotech*. **2007**, *18*, 325102–325112.

14. Schrand, A. M.; Huang, H.; Carlson, C.; Schlager, J. J.; Ōsawa, E.; Hussain, S. M.; Dai. L. *J. Phys. Chem. B* **2007**, *111*, 2–7.

15. Schrand, A. M.; Dai, L. M.; Schlager, J. J.; Hussain, S. M.; Ōsawa, E. *Diamond Relat. Mater.* **2007**, *16*, 2118–2123.

16. Neugart, F.; Zappe, A.; Jelezko, F.; Tietz, C.; Boudou, J. P.; Krueger, A.; Wrachtrup, J. *Nano Lett.* **2007**, *7*, 3588–3591.

17. Gruber, A.; Dräbenstedt, A.; Tietz, C.; Fleury, L.; Wrachtrup, J.; von Borczyskowski, C. *Science*, **1997**, *276*, 2012–2014.

18. Treussart, F.; Jacques, V.; Wu, E.; Gacoin, T.; Grangier, P.; Roch, J.-F. *Physica B*, **2006**, *376*, 926–929.

19. Fu, C.-C.; Lee, H.-Y.; Chen, K.; Lim, T.-S.; Wu, H.-Y.; Lin, P.-K.; Wei, P.-K.; Tsao, P.-H.; Chang, H.-C.; Fann, W. *Proc. Natl. Acad. Sci. USA* **2007**, *104*, 727–732.

20. Barnard, A. S. *Analyst*, **2009**, *134*, 1751–1764.

21. Collins, A. T.; Davies, G.; Kanda, H.; Woods, G. S. *J. Phys. C* **1988**, *21*, 1363–1367.

22. Rabeau, J. R.; Stacey, A.; Rabeau, A.; Prawer, S.; Jelezko, F.; Mirza, I.; Wrachtrup, J. *Nano Lett.* **2007**, *7*, 3433–3437.

23. Chang, Y.-R.; Lee, H.-Y.; Chen, K.; Chang, C.-C.; Tsai, D.-S.; Fu, C.-C.; Lim, T.-S.; Tzeng, Y.-K.; Fang, C.-Y.; Han, C.-C.; Chang, H.-C.; Fann, W. *Nature Nanotech.* **2008**, *3*, 284–288.

24. Mainwood, A. *Phys Rev. B* **1994**, *49*, 7934–7940.

25. Iakoubovskii, K.; Adriaenssens, G. J. *J. Phys.: Condens. Matter,* **2001**, *13*, 6015–6018.

26. Davies, G.; Hamer, M. F. *Proc. R. Soc. Lond. A* **1976**, *348*, 285–298.

27. Davies, G. *J. Phys. C: Solid State Phys*. **1979**, *12*, 2551–2566.

28. Smith, B. R.; Inglis, D.; Sandnes, B.; Rabeau, J. R.; Zvyagin, A. V.; Gruber, D.; Noble, C.; Vogel, R.; Ōsawa, E.; Plakhotnik, T. *Small,* **2009** DOI: 10.1002/smll.200801802

29. Tisler, J.; Balasubramanian, G.; Naydenov, B.; Kolesov, R.; Grotz, B.; Reuter, R.; Boudou, J.-P.; Curmi, P. A.; Sennour, M.; Thorel,A.; Börsch, M.; Aulenbacher, K.; Erdmann, R.; Hemmer, P. R.; Jelezko, F.; Wrachtrup J. *ACS Nano*, **2009,** *3* 1959–1965.

30. Dolmatov, V.Yu. *Rus. Chem. Rev*. **2001**, *70*, 607–626.

31. Danilenko, V.V. *Phys. Solid State* **2004**, *6*, 595–599.

32. Vlasov, I. I.; Barnard, A. S.; Ralchenko, V. G.; Lebedev, O. I.; Kanzyuba, M. V.; Saveliev, A. V.; Konov, V. I.; Goovaerts, E. *Adv. Mater*. **2009**, *21*, 808–812.

33. Sattel, S.; Robertson, J.; Tass, Z.; Scheib, M.; Wiescher, D.; Ehrhardt, H. *Diamond Relat. Mater*. **1997**, *6*, 255–260.

34. Gruen, D. M. *Annu. Rev. Mater. Sci.* **1999**, *29*, 211–259.

35. Sharda, T.; Soga, T.; Jimbo, T.; Umeno, M. *Diamond Relat. Mater*. **2001**, *10,* 1592–1596.

36. Wang, T.; Xin, H. W.; Zhang, Z. M.; Dai, Y. B.; Shen, H. S. *Diamond Relat. Mater*. **2004**, *13*, 6–13.

37. a) Barnard, A. S.; Sternberg, M. *J. Phys Chem B* **2005**, *109*, 17107–17112; b) *ibid. Nanotech.* **2007** *18*, 025702–025713; c) *ibid. Diamond & Relat. Mater*. **2007**, *16*, 2078–2082.

38. Barnard, A. S.; Russo, S. P.; Snook, I. K. *J. Comput. Theo. Nanosci*. **2005**, *2*, 180–201.

39. Turner, S.; Lebedev, O. I.; Shenderova, O.; Vlasov, I. I.; Verbeeck, J.; Van Tendeloo, G. *Adv. Funct. Mater*. **2009**, *19*, 2116–2124.

40. Fang, X. W.; Mao, J. D.; Levin, E. M.; Schmidt-Rohr, K. *J. Am. Chem. Soc.* **2009**, *131*,1426–1435.

41. Shenderova, O. A.; Hu, Z.; Brenner D. in: D. Gruen, A. Ya. Vul', O. Shenderova, Eds., Synthesis, Properties and Applications of Ultrananocrystalline Diamond, Proceedings of the NATO Advanced Research Workshop on Ultrananocrystalline Diamond, St. Petersburg, Russia, NATO Science **2004**.

42. Smith, A. Mainwood, A.; Watkins, M. *Diamond Relat. Mater*. **2002**, *11*, 312–315.

43. Porezag, D.; Frauenheim, Th.; Köhler, Th.; Seifert, G.; Kaschner, R. *Phys. Rev. B* **1995**, *51*, 12947–12957.

44. Frauenheim, Th.; Seifert, G.; Elstner, M.; Niehaus, Th.; Köhler, C.; Amkreutz, M.; Sternberg, M.; Hajnal, Z.; Di Carlo, A.; Suhai, S. *J Phys: Condens. Matter*, **2002**, *14*, 3015–3047.

45. a) Barnard, A. S.; Sternberg, M. *J. Mater. Chem*. **2007**, *17*, 4811–4819; b) Barnard, A.S. *J. Mater. Chem*. **2008**, *18*, 4038–4041

46. Barnard, A. S. *Cryst. Growth Des.*, **2009**, DOI**:** 10.1021/cg90068

47. Beveratos, A. Brouri, R. Gacoin, T. Poizat, J.-P.; Grangier, P. *Phys. Rev. A* **2001**, *64*, 061802–061806.

48. Martin, J.; Wannemacher, R.; Teichert, J.; Bischoff, L.; Köhler. B. *Appl. Phys. Lett*. **1999**, *75*, 3096–3099.

49. Error in the experimental data comes from two main sources: error in the estimate of crystal size using the atomic force microscope and the spherical approximation, and error in the estimation of the number of N-V centres using the contrast in the auto-correlation function. It is expected that these sources of error would lead to no more than 5% error in the final analysis.

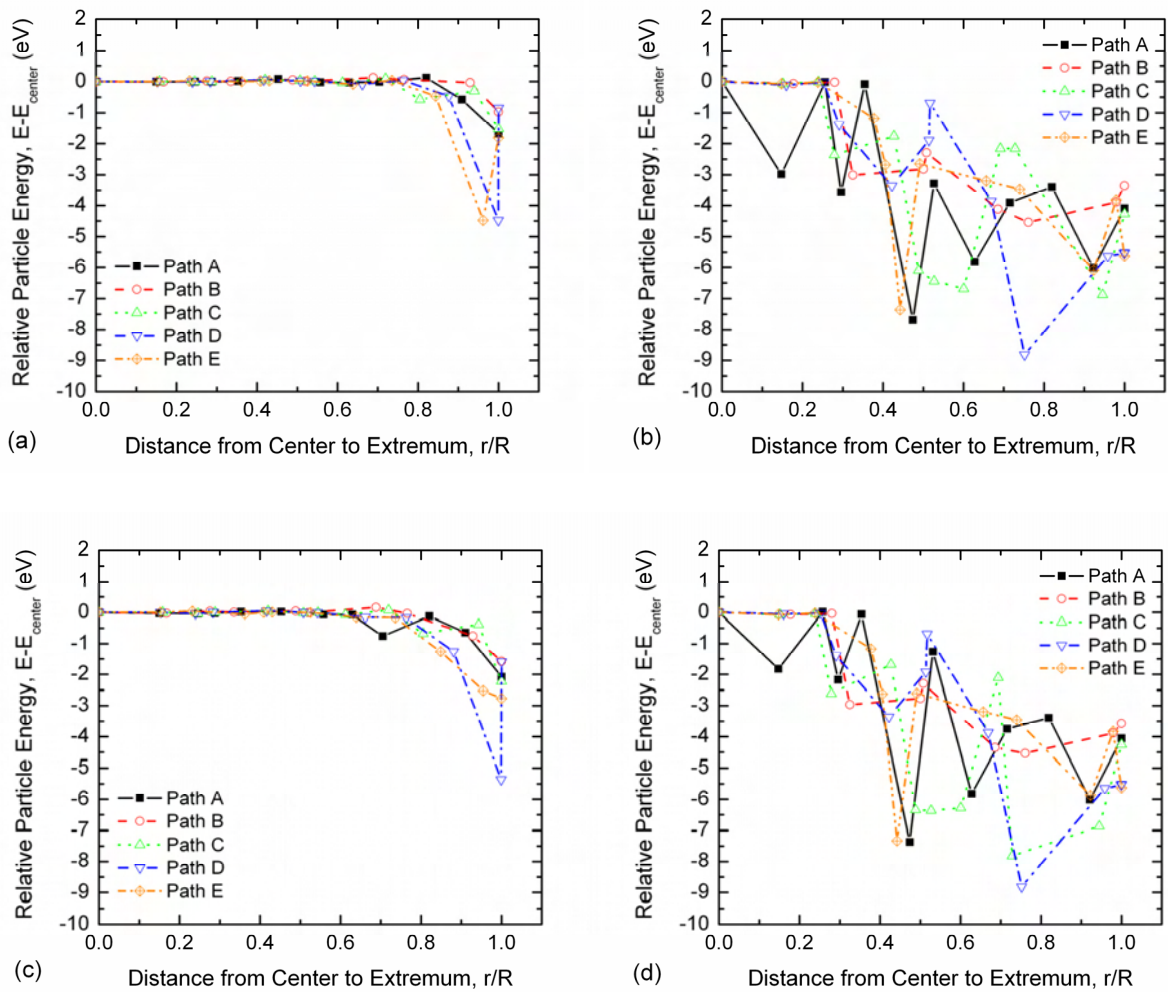

**Figure 1**: Stability of **(a)** $(N–V)^0$ in $C_{837}H_{252}$, **(b)** $(N–V)^0$ in $C_{837}$, (c) $(N–V)^-$ in $C_{837}H_{252}$, (d) $(N–V)^-$ in $C_{837}$. Paths defined in reference 37 (as applied of substitutional N defects).

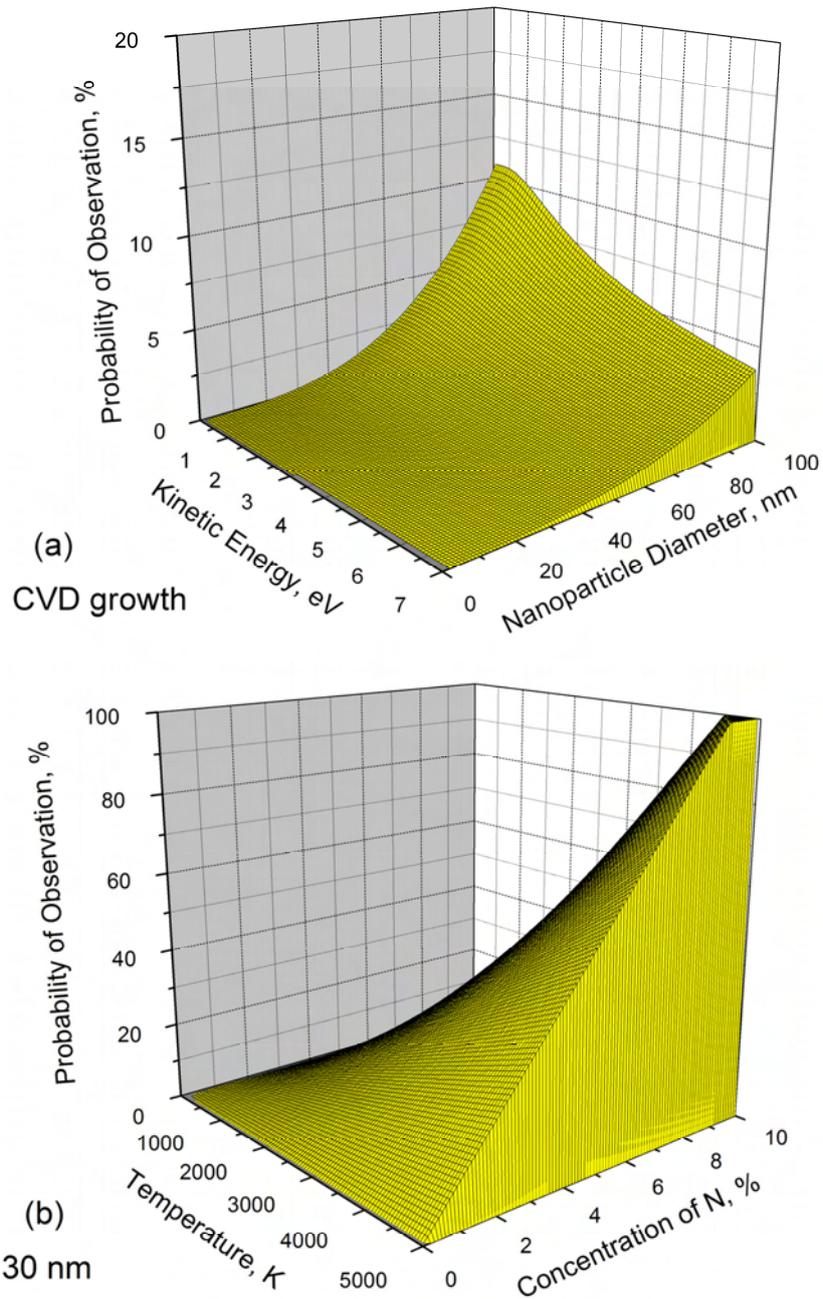

**Figure 2**: Probability of observation of stable N–V defect in diamond nanoparticles: **(a)** over a range of particle diameters and kinetic energies during probing (using CVD growth conditions: $C = 0.1\%$ and $T_{growth} = 800$ K), and **(b)** for 30 nm particles over a range of synthesis temperatures and concentrations of nitrogen present in the precursor materials. Note the difference in vertical scales.

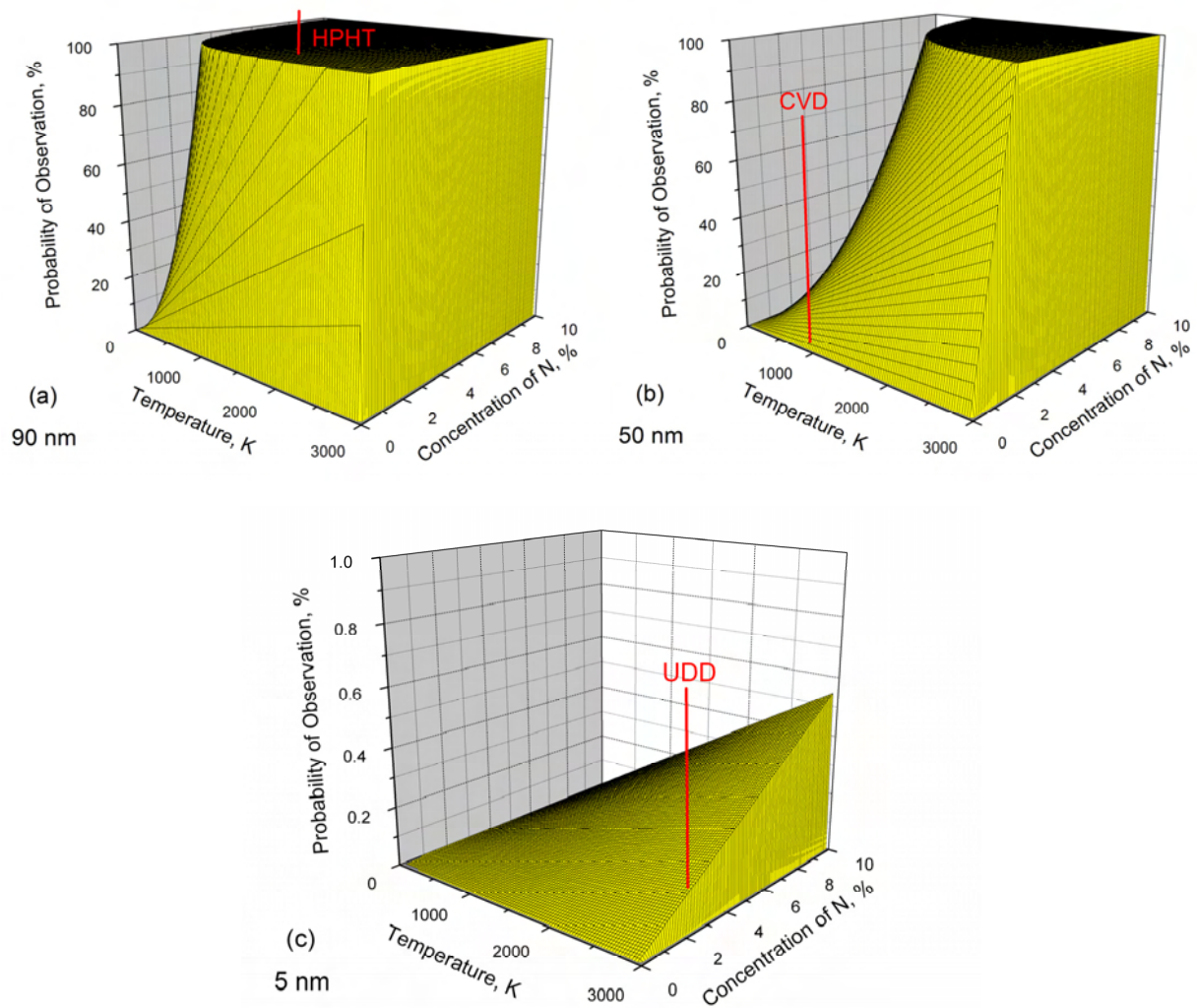

**Figure 3**: Probability of observation of N–V defects in diamond nanoparticles measuring **(a)** 90 nm, **(b)** 50 nm, and **(c)** 5 nm in diameter. The traditional synthesis parameters ($T_{growth}$ and $C$) corresponding to high-pressure high-temperature (HPHT), chemical vapour deposition (CVD) and detonation or ultra-dispersed (UDD) particles are marked. Note the difference in vertical scale in (c).

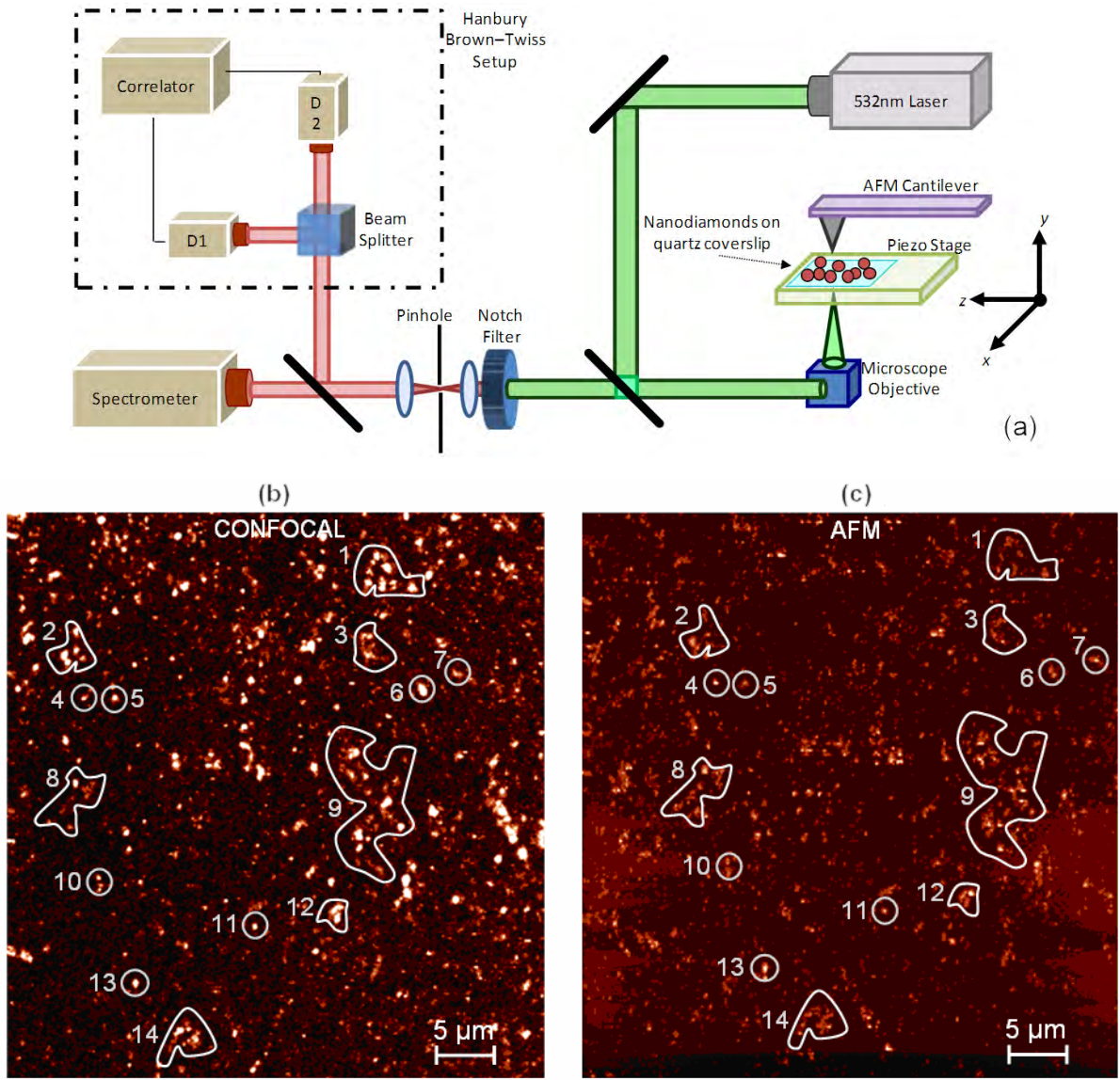

**Figure 4:** **(a)** Combined confocal/AFM system setup **(b)** Confocal system image: bright fluorescing spots indicate emission from N–V centre(s) in nanodiamond crystals. **(c)** Corresponding atomic force microscope (AFM) image of nanocrystalline diamonds deposited on quartz substrate: the brightness of the spots is directly proportional to the height of the crystals themselves.

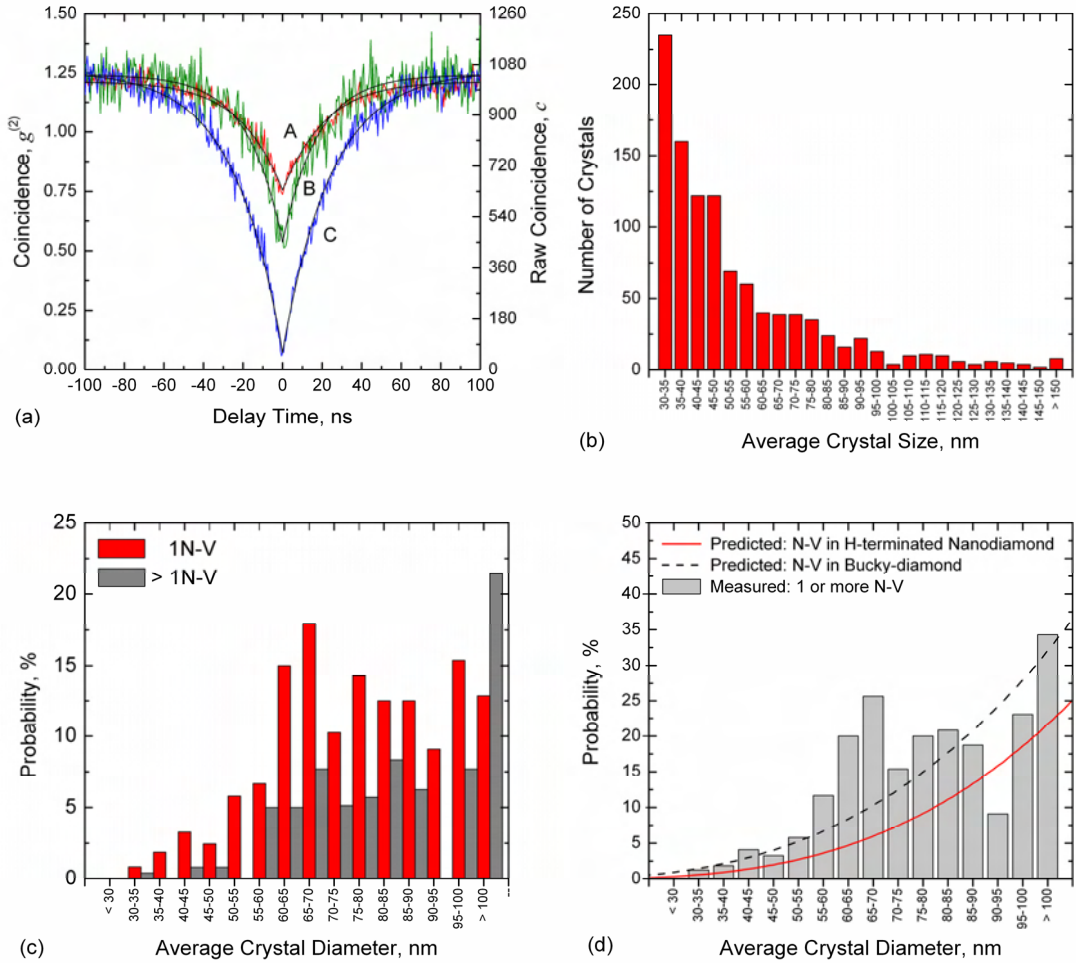

**Figure 5: (a)** Autocorrelation function g(2)(t) (left) and raw coincidence rate c(t) (right) for three different crystals hosting N-V centre(s). **(b)** Histogram of the overall sizes of the diamond nanocrystals. The sample region examined is $50 \times 50$ μm and the total number of crystal found is 3690, with 2624 crystals whose diameter was less than 30 nm (data not shown). **(c)** Experimental measurements of the probability of finding 1 N–V defect (red columns) and the probability of finding more than 1 N–V (grey columns) in HPHT diamond nanoparticles. **(d)** Total probability for a given crystal size range to contain 1 or more N–V colour centre(s), at room temperature. Histogram: experimental measurements combining the probability of finding 1 N–V defect and the probability of finding more than 1 N–V in HPHT diamond nanoparticles. Lines: corresponding theoretically predicted probability, according to equation (5), of finding single N–V defects into diamond nanoparticles for bucky-diamond (dashed) and for H-terminated nanodiamond (solid).

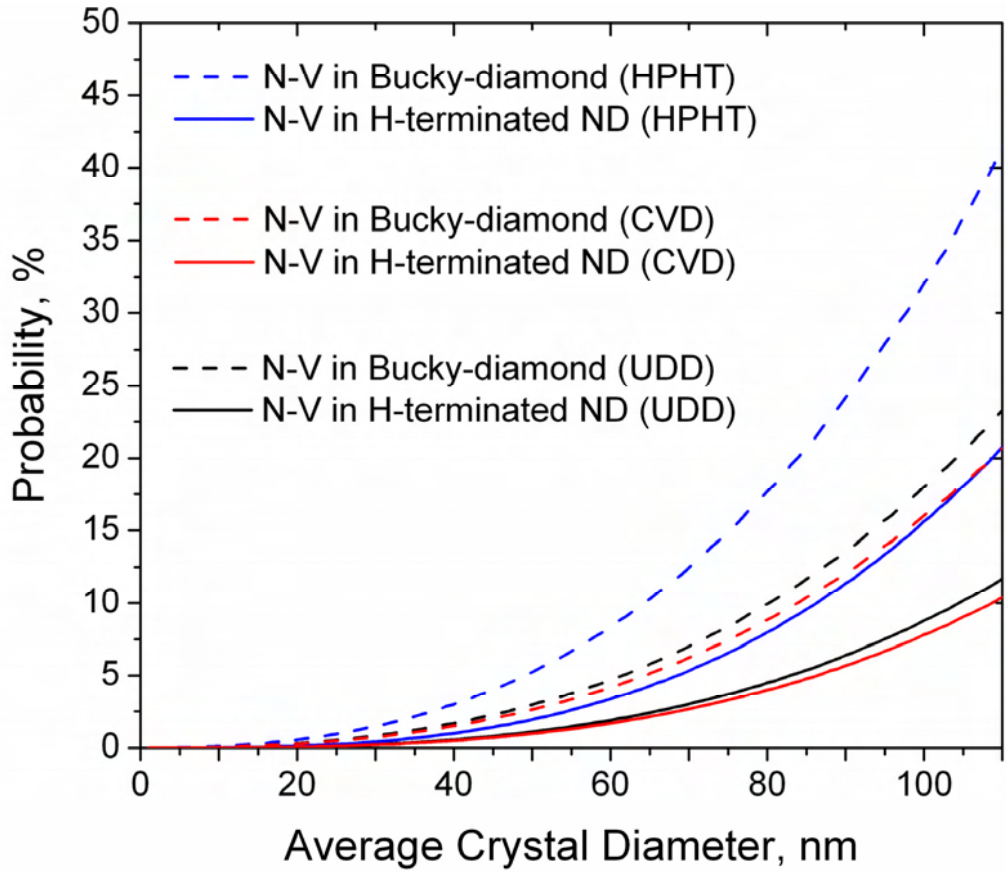

**Figure 6**: Probability of observation of (N–V)⁻ defects in diamond nanoparticles produced using different synthesis methods, as a function of diameter. The traditional synthesis parameters ($T_{growth}$ and $C$) corresponding to high-pressure high-temperature (HPHT), chemical vapour deposition (CVD) and detonation or ultra-dispersed (UDD) particles are used, and a probe kinetic energy of 532 nm has been applied. Comparable results for the neutral (N–V)⁰ defects are indistinguishable within the model limitations.